\documentstyle[preprint,epsfig,aps,prl]{revtex}

\tighten
\begin{document}

\title{Cold Collision Frequency Shift of the 1S-2S Transition in Hydrogen}

\author{Thomas~C.~Killian, Dale~G.~Fried,  Lorenz~Willmann,
David~Landhuis, Stephen~C.~Moss, Thomas~J.~Greytak, and Daniel~Kleppner }

\address{Department of Physics and Center for Materials Science and
Engineering,\\ Massachusetts Institute of Technology, Cambridge,
Massachusetts 02139}

\date{Accepted for publication in Physical Review Letters: October 2, 1998}

\maketitle

\begin{abstract}
We have observed the 
cold collision frequency shift of the $1S$-$2S$ transition in 
trapped spin-polarized
atomic hydrogen. We find 
$\Delta \nu_{1S-2S}=-3.8 \pm 0.8 \times 10^{-10}~n {\rm~Hz~cm}^{3}$, 
where $n$ is the sample density.
From this we derive the $1S$-$2S$ $s$-wave triplet scattering length,
$a_{1S-2S}=-1.4 \pm 0.3$~nm, which is in fair agreement with a recent
calculation.
The shift provides a valuable probe
of the distribution of densities in a trapped sample.

\end{abstract}


Methods for cooling and trapping neutral atoms have opened 
a low energy regime where collisional phenomena
involve only a single partial wave.
The ground state $s$-wave scattering lengths 
for many of the alkali metal  atoms
have
been obtained from combinations of theory and
experiment.
Those studies were motivated in part 
by the need to estimate evaporative cooling
rates\cite{kdr96}, 
study the stability of Bose-Einstein 
condensates\cite{mbg97,bsh97}, 
and understand collisional shifts of 
atomic fountain frequency standards\cite{gc93,kvg97}.

We have measured a
density-dependent frequency shift of the
two-photon $1S$-$2S$ transition in trapped
atomic hydrogen. This allows
us to deduce the $1S$-$2S$ triplet scattering length, 
which has been calculated by Jamieson, Dalgarno, 
and Doyle\cite{jdd96}. 
The frequency shift also
provides a probe of the density distribution
of our trapped sample, 
and constitutes a valuable tool for the study 
of Bose-Einstein condensation. (See the
accompanying paper on BEC in atomic hydrogen\cite{fkw98}.)

Interactions between
neighboring atoms
shift and broaden atomic
spectral lines.
For bosons at low temperatures, where only
$s$-wave collisions occur (below about 1~K for H\cite{jmi89}),
the elastic contribution to the shift can be described
through the density-dependent mean field level shift\cite{pat72}
\begin{equation}
E = {8\pi\hbar^2 a n \over m}.
\end{equation}
Here, $a$ is the atom's $s$-wave scattering length 
for collisions in the gas, and all particles
have mass $m$. The gas has density $n$, and
is assumed to be nondegenerate. 
The difference 
between the mean field energy in the
excited and ground states results in  a frequency shift. 
For weak excitation
of the $1S$-$2S$ transition in
spin-polarized hydrogen,
\begin{equation}
\Delta \nu_{1S-2S} = (a_{1S-2S} - a_{1S-1S}){4\hbar n_{1S} \over m},
\end{equation}
where the scattering lengths are for the triplet potentials.

Additional contributions to the
shift can result from inelastic
processes. 
The most important corrections  should arise from 
collisional hyperfine
transitions and quenching of the $2S$ state. 
Hyperfine transitions in the spin-polarized
sample arise only through weak magnetic dipole
interactions\cite{skv88}, and in our experiment
we observe no evidence for
collisional quenching
on a millisecond time scale. 
Thus it is
reasonable to neglect inelastic processes
when describing the observed
density-dependent frequency shift.

Analogous frequency
shifts are observed in  hydrogen masers \cite{vks87} 
and cesium atomic
fountains \cite{tvs92}. 
Using the collision-based formalism
commonly applied to these systems, one can
derive Eq.\ 2 and also show that
homogeneous broadening due to elastic collisions 
is negligible in the present study
because the phase shift per collision,
$2\pi a/ \lambda_{\rm deBroglie}$, is small.

Our techniques for cooling and trapping hydrogen have
been described elsewhere\cite{gre95}
and are only briefly summarized here.
Molecules are dissociated in a cryogenic discharge, and
the atoms thermalize by collisions with the 250~mK liquid
$^4$He-coated cell wall and each other. A fraction of the   
``low-field seeking'' $F=1, m_F=1$ atoms settle into
the magnetic field minimum of a 500~mK deep
Ioffe-Pritchard  magnetic trap\cite{pri83} (Fig.\ 1).
An axial bias field of a few gauss in the trapping region
inhibits nonadiabatic spin flips.

The wall temperature is decreased to 120~mK. This 
increases the wall residence time of H atoms,
facilitating recombination, 
so that contact with the wall effectively removes energetic
atoms. This process initiates evaporative 
cooling\cite{hes86,mds88} and
thermally disconnects the sample from the wall. 
The evaporation is forced 
by decreasing the trap depth, which is
set by a saddlepoint in the magnetic field.
Sample temperatures as low
as 110 $\mu$K and densities approaching 10$^{14}$ cm$^{-3}$ are achieved.
Under these conditions the sample is a needle-shaped cloud with
a diameter of 200~$\mu$m and 
a length of 8~cm.

The speed with which the
trap threshold can be lowered is limited 
by the need to maintain thermal 
equilibrium through elastic collisions during the evaporation.
After the forced evaporation is terminated, 
the sample temperature remains constant.
Its value 
is determined by the balance between cooling  due
to evaporation of atoms over the threshold and heating due to 
dipolar spin-flip collisions. Dipolar decay
preferentially removes atoms from the bottom of the trap
where the density is highest.

The 
temperature\cite{dsm89} and density\cite{mds88} are measured
by rapidly lowering the trap threshold
to zero while monitoring the power deposited on a bolometer.
The power results from molecular recombination and
is proportional to the flux of atoms escaping from the trap.
The dependence of the flux on the  threshold reveals
the energy distribution of atoms in the trap, which,
combined with a knowledge of the trapping
fields, yields the sample temperature.

We measure the density 
by releasing the atoms from the trap after holding them
for different times $t$ following 
the forced evaporation.
The total
recombination energy deposited on the bolometer 
is proportional
to $N(t)$, the number of atoms remaining in the trap
at time $t$.
$N(t)$ decreases with time 
due to dipolar decay, which
removes atoms from the sample at a 
density-dependent rate (Fig.\ 2). 
This process obeys the
two-body equation 
${\rm d} n /{\rm d}t = -g n^2$, 
where $n$ is the local density
and $g=1.1 \times 10^{-15}$ cm$^3$/s is the calculated
rate constant\cite{skv88,gnote}. 
When averaged over the density distribution in the trap, 
this implies that  $N(t)$ obeys\cite{mds88} 
\begin{equation}
N(0)/N(t) = 1 + \kappa g n_0(0) t.
\end{equation}
Here, $n_0(t)$ is the maximum density in the sample
at time $t$,
and $\kappa$ is 
a numerical factor 
which depends on the trap geometry.
Computations show that $\kappa$ is typically 
in the range $0.2$--$0.3$. For a given geometry,
the uncertainty in $\kappa$ is
10\%, arising
from imperfect knowledge of
the magnetic fields.  
The number of atoms for a  given trapping configuration is
reproducible to a few percent.

Once the sample temperature and density are known,
we measure the cold collision
frequency shift by two-photon Doppler-free
$1S$-$2S$ spectroscopy\cite{cfk96}.
The 243 nm laser that excites atoms to the
metastable $2S$ state is
stabilized by an optical cavity and has
a linewidth of about 1~kHz.
8~mW of radiation 
is focused to a 
waist radius of $w_{0}=50~\mu$m in the trap. 
The beam lies along the axis of the
atom cloud and is retro-reflected by a spherical mirror to
provide the standing wave required for Doppler-free
excitation.  
A mechanical chopper pulses the laser beam at
1~kHz with a 50\% duty cycle. 
Following each pulse, an electric field of $10~{\rm V/cm}$ 
Stark-quenches the excited atoms 
by mixing the $2S$ and $2P$ wavefunctions. This causes
prompt radiative decay
by emission of a Lyman-$\alpha$ photon ($122$~nm).  The emitted photons are
counted using a microchannel plate detector located at the end of the
trap, behind the retro-reflecting mirror.  Due to the low collection
solid angle ($2 \times 10^{-2}$~ster), 
absorptive losses, and detector quantum efficiency, 
the total detection efficiency is only $10^{-5}$.  

Typical data for determining the 
cold collision frequency shift is shown in
Fig.\ 3a.
The density is highest immediately after the forced evaporation,
so the first recorded spectrum is shifted the most.
Due to the distribution of
densities in the sample, the line is 
inhomogeneously broadened.
The density decreases on a 20 s time scale primarily
due to collisions 
with helium atoms evaporated
from the mirror surface by the laser; hence
successive scans probe different densities.
The later spectra exhibit 
the exponential lineshape, 
exp$(-|\nu | /\delta\nu_{\rm transit})$ \cite{bbc79},
expected for Doppler-free two-photon excitation 
by a Gaussian laser beam of 
a low density sample.
Here, $\nu$ is the laser detuning from resonance at 243 nm in Hz,
$\delta \nu_{transit}=\sqrt{2k_BT/m}/(4\pi w_0)$ 
is the linewidth due to 
the finite interaction
time of an atom with the laser beam~--~typically 2--5~kHz. 
Measurement of the low density linewidth 
provides an independent check of our sample temperature
\cite{cfk96}.

To quantify the cold collision frequency shift, one must account 
for the distribution of densities in the trap. 
Numerical simulations of the spectra show that the shift of the 
line center is about 0.75 of the shift associated with the maximum
density in the sample, $n_0$.
This factor depends on
the trapping magnetic fields, sample temperature,
and laser focus position with respect to the atom cloud.
There is a 10\% uncertainty in this correction for a given
configuration.

The integral of the $1S$-$2S$ spectrum is proportional to 
$n_0(t)$, and it decreases smoothly in time.
This allows us to extrapolate from the initial
density before the first scan, found 
as described above, to determine the density for
each successive scan. 
A fit to data such as shown in Fig.\ 3b
yields $\chi$, where $\chi=\Delta \nu_{1S-2S} / n$.
Measurement errors are small compared to 
the mentioned systematic uncertainties.
The long term drift of the laser frequency,
$\sim$ 10 kHz/hour, is
taken into account. Note that no
knowledge of absolute frequency is required to determine 
$\chi$ from a given series of scans.

To examine the sensitivity to systematic effects, 
$\chi$ was measured in
different trap configurations,
with sample temperatures between 110 and 500~$\mu$K 
and initial maximum densities in the range 
$(2\sim7) \times 10^{13}$~cm$^{-3}$.
In Fig.\ 4, the various measurements are plotted
versus sample temperature.
We observe no significant 
temperature dependence of the shift,
consistent with theory.
Any significant nonlinear density dependence of the
shift would manifest itself
in data such as Fig.\ 3b,
but none is evident.

From a weighted average of the various measurements, we find
$\chi=-3.8 \pm 0.8 \times 10^{-10} {\rm~Hz~cm}^3$.
In an apparatus optimized for spectroscopy 
with better known magnetic
field and laser geometries, 
the uncertainty could be greatly reduced.
A revision of the 
calculated value for $g$ would cause
a proportional change in the value of $\chi$.
Experiments have verified
this decay constant at the 20\% level\cite{rbj88},
but since the calculation
involves well known ground state potentials, the 
calculated value is expected
to be more certain. 

Equation 2 relates $\chi$ to the $s$-wave 
triplet scattering lengths,  
assuming inelastic processes can be neglected.
From this, we derive
$a_{1S-2S}=-1.4\pm0.3$~nm.
We have used the theoretical value of 
$a_{1S-1S} = 0.0648$~nm\cite{jdk95},
which constitutes only a small 
contribution to $\chi$.
The value calculated by Jamieson {\it et al.}\cite{jdd96}, 
$a_{1S-2S} = -2.3$~nm, 
is in fair agreement with this measurement.

Having calibrated the cold collision frequency shift, 
we can use two-photon spectroscopy to measure
sample density provided the shift is comparable to or greater than the
transit-time width, $\delta \nu_{\rm transit}$. 
The spectrum, $S(\nu)$, is
given by the approximate expression 
\begin{equation}
S(\nu) \propto \int {\rm d^3r}\ n({\bf r}) 
I^2({\bf r})\delta(\nu - \chi n({\bf r})),
\end{equation}
which neglects  atomic motion and 
laser linewidth.
$I({\bf r})$ is the laser intensity, and
the delta function enforces the resonance
condition.
In our experiment, at high densities and low temperatures, 
and especially in the regime
of Bose-Einstein condensation, 
this probe is more sensitive and accurate than
the bolometer. In addition, it provides valuable
information on the density distribution\cite{fkw98}.

$1S$-$2S$ spectroscopy of cold trapped hydrogen can 
be used
for optical frequency metrology and precision
measurements.
Linewidths of about 1 kHz have been
achieved in an atomic beam\cite{uhg97},
but with the long coherence time possible in a trap\cite{cfk96} 
it should be feasible to 
approach the 1.3 Hz natural
linewidth of the transition.
To reduce the effects of 
the cold collision shift below this level,
it is desirable to work at densities of
$\leq 10^{10}$~cm$^{-3}$. 
Even with such a decreased density,
large signal rates are still possible
in an apparatus optimized for optical access.

We thank Jon C. Sandberg and Claudio L. Cesar for construction
of the laser system used in this experiment and 
development of two-photon
spectroscopy of trapped hydrogen.
We are also grateful to Alexander Dalgarno for helpful discussions.

This work is funded by the National Science Foundation 
and the Office of Naval Research. Early support came
from the Air Force Office of Scientific Research.
L.W. acknowledges support by
Deutsche Forschungsgemeinschaft. D.L. and S.C.M. are supported by 
National Defense Science and Engineering Graduate
Fellowships.


\begin{figure}
\vspace{0.5in}
\centering\epsfig{file=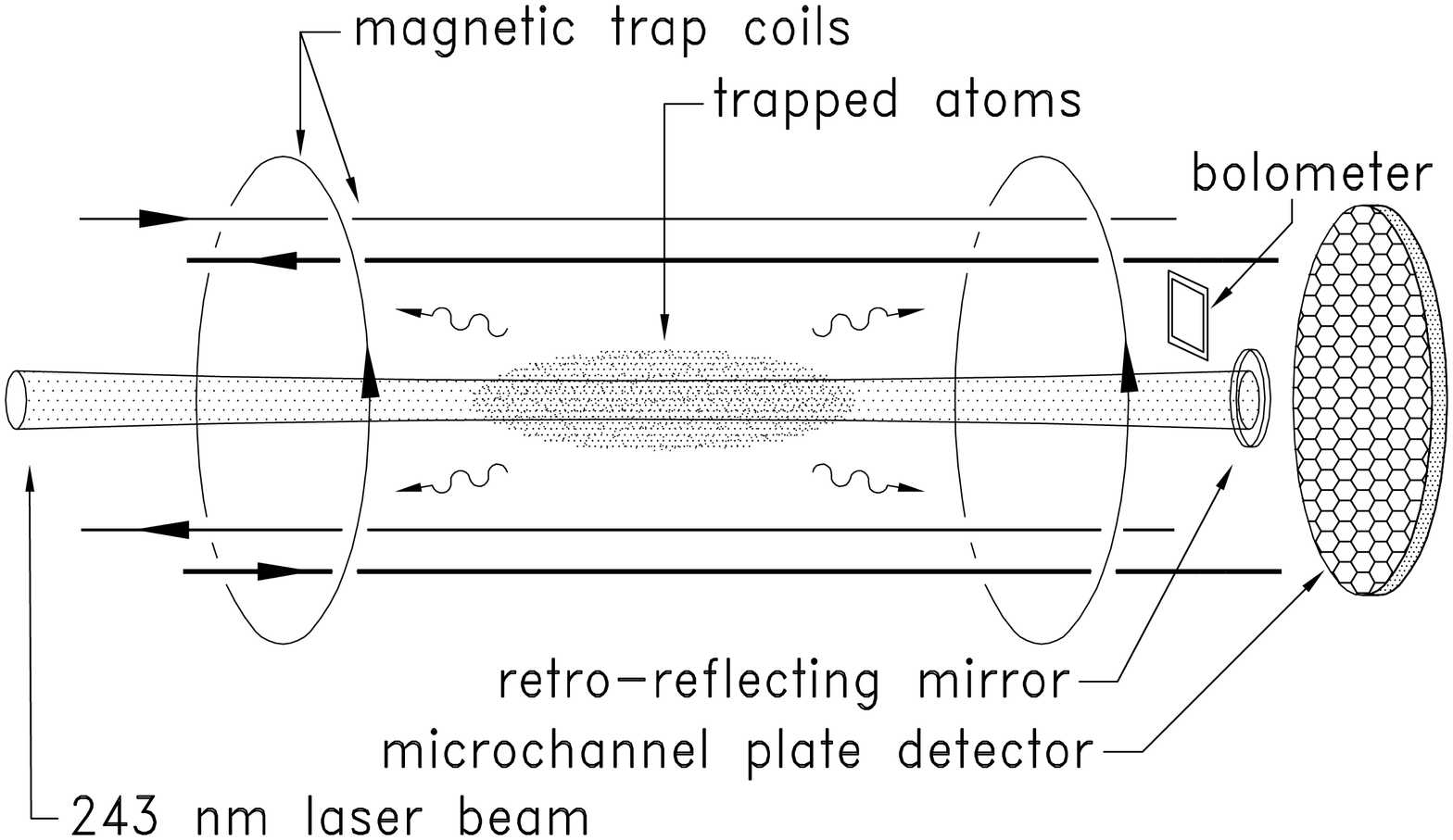, angle=90,width=4in}
\vspace{.125in}
\caption{Schematic diagram of the apparatus.
Coils create a magnetic field with a minimum along
the trap axis, which confines the sample. The 243 nm laser
beam is focused to a 50~$\mu$m beam radius 
and retroreflected. A $500~\mu$s laser pulse promotes some atoms to
the metastable $2S$ state. An electric field
then Stark-quenches the $2S$ atoms, and
the resulting Lyman-$\alpha$ fluorescence photons
are counted by the microchannel
plate. Not shown is the trapping cell which surrounds the 
sample and is thermally anchored to a dilution refrigerator. 
The actual
trap is longer and narrower than indicated in the diagram.} 
\label{apparatus}
\end{figure}

\begin{figure}
\vspace{0.5in}
\centering\epsfig{file=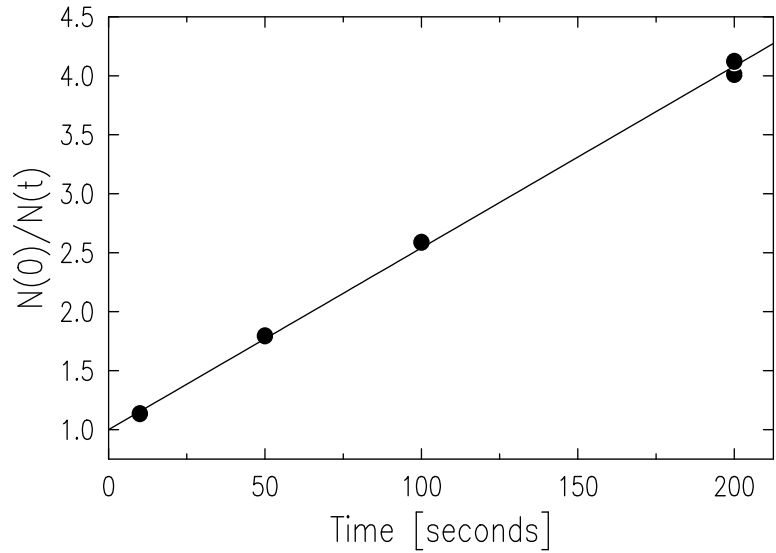, width=4in}
\vspace{.125in}
\caption{Dipolar decay of trapped hydrogen.
The sample density is found from the slope of $N(0)/N(t)$,
the inverse
of the normalized total number of atoms remaining in the trap.
The data shown indicates a density of 
$6.0 \times 10^{13}$~cm$^{-3}$.}
\label{density}
\end{figure}

\begin{figure}
\begin{center}
\epsfig{file=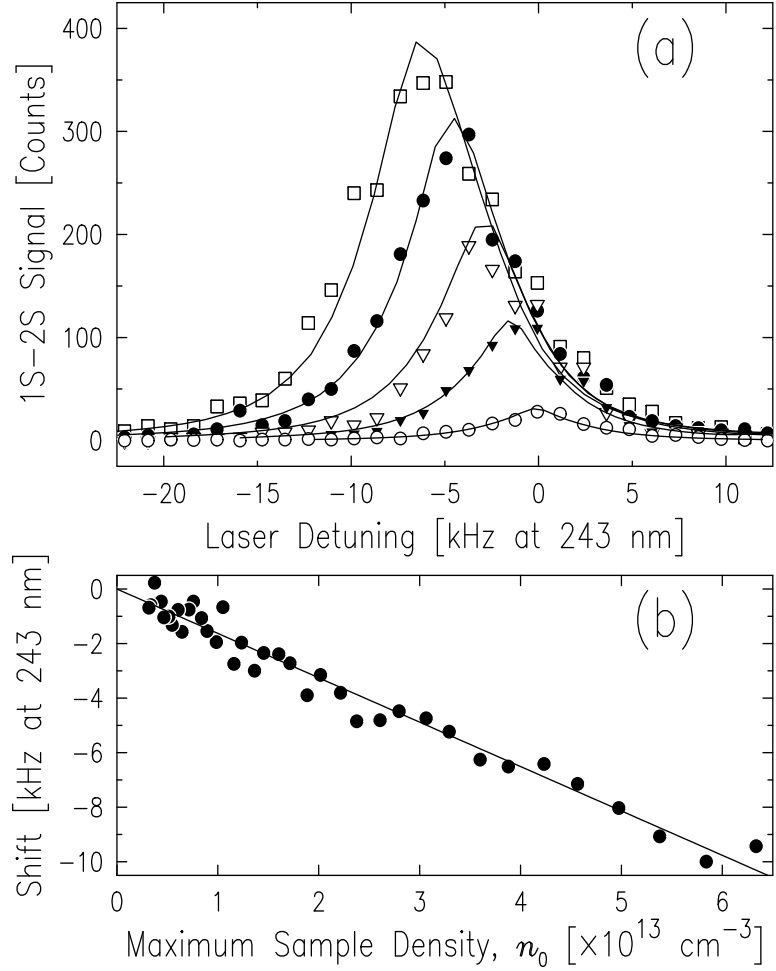, width=4in}
\end{center}
\vspace{.125in}
\caption{(a) Cold collision frequency shift observed in 
the spectra of a single 120~$\mu$K sample with initial maximum
density of $n_0(0) = 6.6\times 10^{13}$~cm$^{-3}$. 
The furthest red-shifted spectrum was recorded first, 
and subsequent spectra were taken at decreasing sample densities.
Each data point represents
12 ms of laser excitation and one scan is recorded in 0.5 seconds.
20 seconds elapsed between the first and last scan shown. 
The smooth curves are the result of numerical simulation.
Only 5 of the 40 spectra taken of
this sample are shown.\\
(b) Frequency shifts from spectra of
sample described in
(a) after correcting for inhomogeneous sample density. 
For this data, the linear fit yields 
$\chi = -3.30 \pm 0.6 \times 10^{-10} {\rm~Hz~cm}^3$,
where $\chi= \Delta \nu_{1S-2S} / n$. (Note,
$\Delta \nu_{1S-2S}$, the shift in the transition
frequency, is twice the shift in the 
frequency of the 243 nm laser.)
}
\label{shiftingline}
\end{figure}


\begin{figure}
\centering\epsfig{file=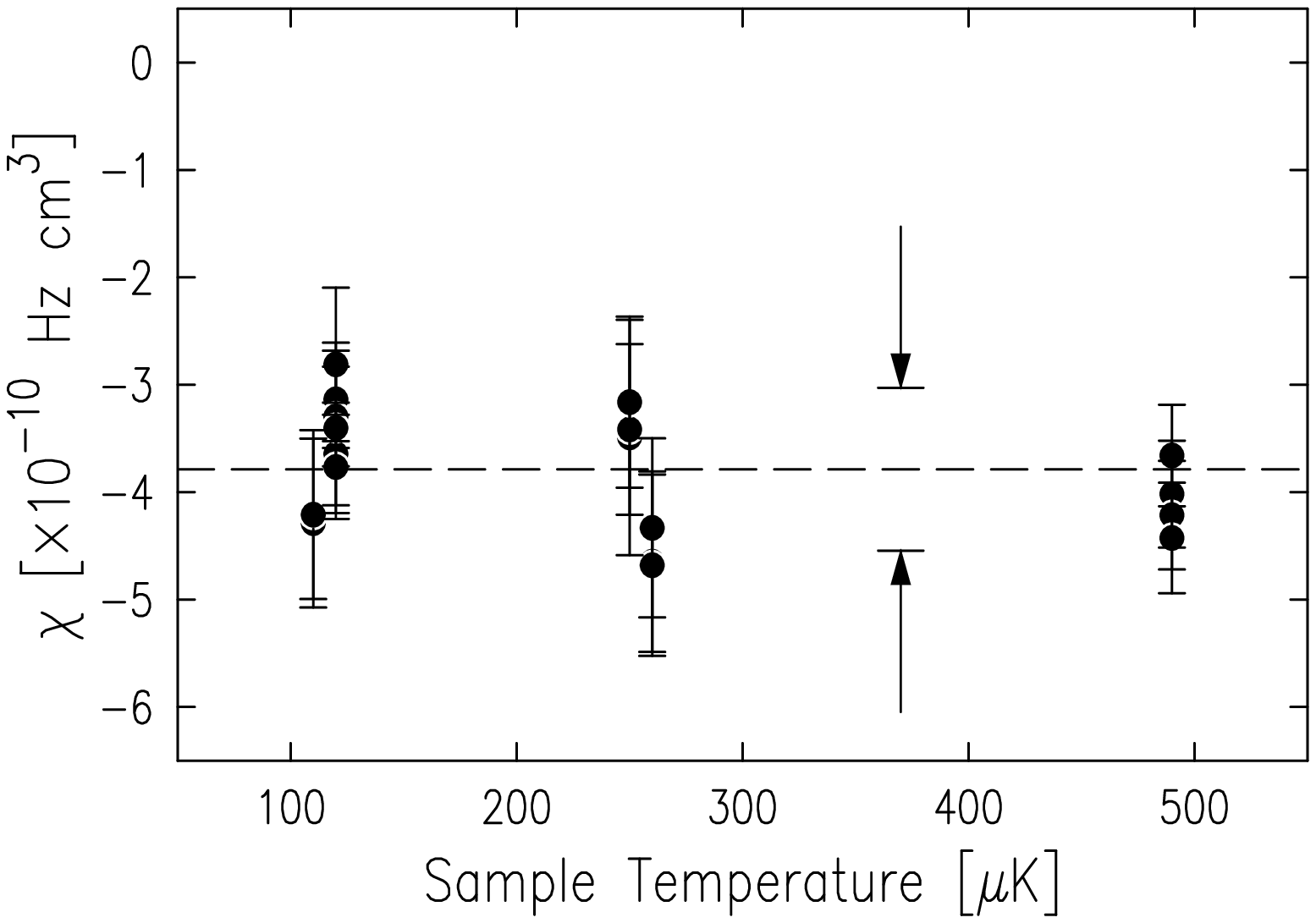, width=4in}
\vspace{.125in}
\caption{The frequency shift parameter $\chi$,
determined
as described 
in Fig.\ 3 for various trap 
configurations. No significant
dependence on sample temperature is observed.
The error bars reflect systematic uncertainties
in the magnetic trapping fields and laser geometry, with small
contributions from measurement error.
The dashed line is the weighted mean
of all measurements, 
$\chi = -3.8 \pm 0.8 \times 10^{-10} {\rm~Hz~cm}^3$,
and the quoted uncertainty is indicated by the double arrows.}
\label{fitshift}
\end{figure}


\begin{references}
\bibitem{gnote} We calculate the rate constant 
$g = 2(G_{dd \rightarrow ac} + G_{dd \rightarrow aa} + 
G_{dd \rightarrow ad})$,
using zero temperature and zero field values of
the decay event rates, $G$.
Atoms ending 
in the $c$ and $d$ states after a spin flip
are lost from the trap because the
hyperfine energy (71~mK) is much greater than the trap depths
used in this study.
\end{references}
\end{document}